\documentclass[conference]{IEEEtran}
\usepackage{amsmath,amssymb,amsthm}
\usepackage{dblfloatfix}
\usepackage[english]{babel}
\usepackage{graphicx}
\usepackage{cite}
\usepackage{algorithm}
\usepackage{algorithmic}
\usepackage{listings}
\usepackage[table,xcdraw]{xcolor}

\begin{document}
\title{Hardware-In-The-Loop Vulnerability Analysis of a Single-Machine Infinite-Bus Power System}
\author{\IEEEauthorblockN{Hossein Salehghaffari }
\IEEEauthorblockA{Control/Robotics Research Laboratory (CRRL),\\
Department of Electrical and Computer Engineering,\\ NYU Tandon School of Engineering (Polytechnic Institute), NY 11201, USA \\
Email: \{h.saleh\}@nyu.edu
}}
\maketitle

\begin{abstract}
The dynamic performance of the generators is a critical factor for the safe operation of the power grid. To this extent, the stability of the frequency of generators is the target of cyber attacks since its instability may lead to sizable cascade failures in the whole network. In this paper, we perform the vulnerability analysis in a developed power grid Hardware-In-The-Loop (HITL) testbed with a Wago 750-881 PLC sending control commands to the generators and a 750 Feeder Management Relay connected to a local load. A process-aware coordinated attack is demonstrated by spoofing control commands sent by the PLC and the relay to the simulated power system which is modeled as a single-machine infinite-bus (SMIB). Based on the reachability analysis, the attacker can find the optimal attack signal to drive the system state out of their safe set of values. Thereafter, it is experimentally demonstrated that the attacker does not need to send attack signal continuously if he implements a carefully designed coordinated attack on the PLC and the relay. The presented assessments provide information about the best time to launch an attack in order to destabilize the power system.

\end{abstract}
\vspace*{0.1cm}
\section{Introduction}

There have been several attacks on industrial control systems such as power grids, chemical processes, and water infrastructures in recent years\cite{khorrami2016,khodaparastan20121,salehghaffari2017}. Due to several economical and technological factors, the integration of IT infrastructures that support the timely exchange of data into the electrical grids is growing fast. Also, today's power electric systems are equipped with different power storage technologies \cite{khodaparastan2017study,super} that may lead to different security concerns.  Consequently, power grids dependency on Supervisory Control and Data Acquisition (SCADA) have made them vulnerable to various types of cyber attacks. Development of reliable large power systems requires profound understating about the potential cyber attacks on IT infrastructures and their impacts on the performance of the power systems.

There have been substantial works on different types of attacks on SCADA system of power grids\cite{kosut2011,teixeira2010,kocc2014,chen2010,giani2013}. different detection and control methods including hybrid \cite{khodaparastan2017novel} and continuous time \cite{khodaparastan2012} are utilized in power grids.  In false data injection (FDI) attacks, the attacker manipulates the data generated by the SCADA system to deteriorate the performance of state estimation unit. In \cite{teixeira2010}, the problem of cyber security of the state estimator in SCADA system is investigated and the paper proposes two metrics to quantify the impacts of possible attacks on the different level of model uncertainties. \cite{kosut2011} introduces the strong attack regime concept and examines the most damaging attack as the adversary has control over several meter measurements. A novel defense strategy based on applying known perturbation on power grid topology is considered in \cite{salehghaffari2018resilient}, and the optimal attack and defense strategies are modeled as a zero-sum game.  \cite{liu2013} considers the effect of denial-of-service (DoS) attacks on communication infrastructures of smart grids, and shows the dynamic performance of load frequency control for a two-area power system in presence of attacks. In \cite{yuan2011}, the load redistribution (LR) attacks as a special type of FDI attacks is introduced, and the impacts of the LR attack on the system performance is analyzed through a game theoretic approach. \cite{li2016} proposes a coordinated optimal cyber attack and analyzes the most damaging attack for the attacker with limited sources.

Moreover, Multiple approaches have been considered to assess the vulnerability of power grids to typical types of attacks\cite{ma2013,sridhar2012,hug2012,correa2013,zhu2013}. \cite{ma2013} analyzes the possible vulnerabilities of transmission lines of power networks against possible attacks while the interaction between the attacker and the defender is captured as a zero-sum stochastic game. In \cite{pasqualetti2011}, a framework is proposed to examine vulnerabilities of different power system topologies to FDI attacks and replay attacks. Thereafter, the attacks identification problem is modeled as an unknown input observer design problem.  In \cite{sridhar2012}, a layered approach is proposed to assess the risk of dependency of cyber security infrastructures and stability of the power systems. Reachability analysis is an effective tool for the verification and the assessment of hybrid systems \cite{esfahani2010,tomlin2003,jin2005,koo2009,lygeros2004}. \cite{tomlin2003} proposes a zero-sum two players game model to assess the aircraft collision avoidance problem and solves the game numerically by leveraging level set methods. In \cite{jin2005}, a numerical method is proposed to compute the stability region of the power systems and the numerical results are utilized to design the controller for the transient stability of the power systems.\cite{lygeros2004} shows that the reachability problem can be formulated as an optimal control problem, and establishes links between reachability, viability, and invariance problems with the solution to a particular Hamilton-Jacobi equation.

In this paper, an integrated time-synchronized HITL simulator with a PLC and a relay is developed. Motivated by \cite{lygeros2004} and utilizing level set toolbox \cite{mitchell2000}, we perform a reachability analysis for the SMIB model connected to the PLC and the relay to obtain deep understanding about the robustness of the power systems with respect to coordinated process-aware attacks on the physical hardware.  The effects of the attack on PLC and relay on the power system performance and stability are experimentally demonstrated, and the optimal attack policy to deteriorate stability of the system is obtained through the reachability analysis. In summary, the main contribution of this paper is threefold:
\begin{itemize}
\item Development of an integrated time-synchronized HITL
simulator for a SMIB connected to a PLC and a relay for studying security related
risks.
\item Experimental demonstration of the effect of PLC and relay vulnerabilities
on the SMIB performance and stability.
\item Development of a systematic approach to assess the robustness of the system to different levels of an attack signal.   
\end{itemize}

This paper is organized as follows: In Section \ref{sec1}, the reachability analysis preliminaries and the mathematical model of SMIB are explained. The structure of the HITL testbed is presented in Section \ref{sec2}. In Section \ref{sec3}, the simulation results related to the stability region computation and the system performance under attacks are demonstrated.   The impact of the attacks on hardware components of the testbed is illustrated in Section \ref{sec4}. In the last section, final remarks and conclusion are presented.

\vspace*{0.1cm}
\section{Problem Formulation} \label{sec1}

\begin{figure}[!t]
\centering
\includegraphics[width=0.30\textwidth]{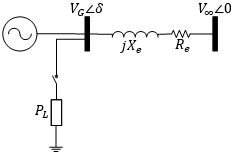}
\caption{Schematics of SMBI model.}
\label{fig0}
\end{figure}

Reachability analysis is an effective tool to examine the vulnerabilities of the power systems whose behaviors are modeled as hybrid mathematical models. The mathematical model of SMIB has been utilized to determine the impact of possible cyber attacks on transient stability of the power systems. The linear swing equation given in (\ref{eq1}) is the mathematical model of SMIB shown in figure \ref{fig0}.

\begin{eqnarray}
\dot{\delta}&=&\omega \nonumber\\
M\dot{\omega}&=&P_m-D\omega-P_E \sin(\delta)-P_L+d \label{eq1}
\end{eqnarray}
where $\delta$ is the generator rotor angle, $\omega$ is the rotor angular velocity deviation, $M$ is the inertia of the rotor, $D$ is the damping coefficient, $P_m$ is the mechanical power input, $P_L$ is the local load connected to the system, $P_E$ is the maximum transfered electrical power, and $d$ is the external disturbance or the attack signal.

The transient stability of the power system frequency is a vital part of the stability of the whole network. Instability of the frequency of SMIB can jeopardize the stability of the whole network and damage the power devices. Therefore, the generator rotor angle and the rotor frequency deviation should be limited to satisfy the safety constraints. The SMIB is considered safe if its variables $\delta$ and $\omega$ satisfy the given inequalities as follows:

\begin{eqnarray}
\frac{-\pi}{2} \leq &\delta -\delta_n& \leq \frac{\pi}{2} \nonumber\\
-6 \leq &\omega-\omega_n& \leq 6 \label{eq0}
\end{eqnarray}
where $\delta_n$ and $\omega_n$ denote the nominal generator rotor angle and the nominal rotor frequency deviation.

Reachability analysis is a powerful numerical tool that provides us with effective information about how long it will take for the system to leave the safe region under a bounded attack signal, what is the stability region of the system, and how much the safe set changes after a specific amount of the time. In our Hardware in the loop simulator, the PLC sends control signals to the actuators, e.g., generation set-points to the generator that mandate how much power the generator should generate and a relay is used to control the connection of the local load $P_L$ to the SMIB model defined in (\ref{eq1}). Also, the system dynamics is simulated by the Matlab software. 

The SMIB dynamics can be written as a continuous nonlinear system $\dot{x}=f(x,d)$ with $x \in R^2$ and $d \in D\subset R$. It is assumed that $D\subset R$ is compact and $f(.)$ is bounded and Lipschitz continuous. Moreover, for any $T\geq 0$, $x \in R^2$, $t \in [0, T]$, and $d \in D$, the system \ref{eq1} admits a unique solution $\phi(.,t.x,d)$ and $\phi(t,t.x.d)=x$.   

Given (\ref{eq1}), we would like to characterize the relation between the set of states $S \in R^2$ and state trajectory of the SMIB over time horizon $[t, T], T\geq t$. The main goal is to answer the following questions by leveraging the reachability analysis: (i) what is the set of the initial conditions that will converge to the stable equilibrium of the system? (ii) what is the set of the initial conditions that the system does not leave the safe set for any attack signal $d \in D$? (iii) what is the set of initial conditions that there exists at least one attack signal $d \in D$ to keep the system in a predefined set? To answer these questions, the following sets should be computed:

\begin{equation}
\begin{split}
Reach(t,S)&=\{x \in R^2| \exists d(.) \in D, \exists\tau \in [t, T], \\
& \phi(\tau, t, x, d(.))\in S\} 
\end{split} \nonumber
\end{equation}
\begin{equation}
\begin{split}
Inv(t,S)&=\{x \in R^2| \forall d(.) \in D, \forall\tau \in [t, T], \\
&\phi(\tau, t, x, d(.))\in S\} 
\end{split} \nonumber
\end{equation}

\begin{equation}
\begin{split}
Viab(t,S)&=\{x \in R^2| \exists d(.) \in D, \forall\tau \in [t, T], \\
&\phi(\tau, t, x, d(.))\in S\} 
\end{split} \nonumber
\end{equation}

Given the set $S=\{x \in R^2| l(x) \geq 0\}$, where $l(x)$ is the sign distance to the set $S$, $Inv(t,S)$ and $Viab(t,S)$ are computed as follows \cite{lygeros2004}:
$Inv(t,S)=\{x \in R^2| V_1(x,t) \geq 0\}$, where the value function $V_1(x,t)$ is given as $V_1(x,t)=\inf_{d(.)\in D} \min_{\tau\in [t, T]}l(\phi(\tau, t, x, d(.)))$. Moreover, $V_1(x,t)$ is the solution of the terminal value problem.
\begin{equation}
\frac{\partial V_1}{\partial t}(x,t) + \min\{0, \inf_{d(.)\in D}\frac{\partial V_1}{\partial x}(x,t)f(x,d)\} \label{eq2}
\end{equation}
with terminal condition $V_1(x,T)=l(x)$. Similar to invariant set computation,  $Viab(t,S)=\{x \in R^2| V_2(x,t) \geq 0\}$, where the value function $V_2(x,t)$ is given as $V_2(x,t)=\sup_{d(.)\in D} \min_{\tau\in [t, T]}l(\phi(\tau, t, x, d(.)))$. Moreover, $V_2(x,t)$ is the solution of the terminal value problem.
\begin{equation}
\frac{\partial V_2}{\partial t}(x,t) + \min\{0, \sup_{d(.)\in D}\frac{\partial V_2}{\partial x}(x,t)f(x,d)\} \label{eq3}
\end{equation}
with terminal condition $V_2(x,T)=l(x)$.

given the definition of $Reach(t,S)$ and $Inv(t,S)$, it is clear that $Reach(t,S)=(Inv(t,S^c))^c$. Therefore, it is not needed to compute $Reach(t,S)$ separately. 

If the nonlinear system $\dot{x}=f(x,d)$ can be written as:
\begin{equation}
\dot{x}=f_x(x)+f_d(x)d
\end{equation}
Then, the optimal input for the equations (\ref{eq2}) and (\ref{eq3}) can be derived for invariant set and viability set respectively as:
\begin{equation}
    d_1^* =\begin{cases} \label{eq4}
      d_h & if \quad \frac{\partial V_1}{\partial x}(x,t)f_d(x) \leq 0, \\
      d_l & \quad \quad \quad otherwise \\
   \end{cases}
\end{equation}
,
\begin{equation}
    d_2^* =\begin{cases} \label{eq5}
      d_l & if \quad \frac{\partial V_2}{\partial x}(x,t)f_d(x) \leq 0, \\
      d_h & \quad \quad \quad otherwise \\
   \end{cases}
\end{equation}
where $d_l$ and $d_h$ are lower and upper limits of the signal $d$, $d_1^*$ is the optimal input seeking to keep the system state out of the safe set and $d_2^*$ is the optimal input seeking to keep the system state in safe set. 

\vspace*{0.1cm}
\section{Experimental HITL Testbed} \label{sec2}
Software models cannot accurately recreate real-world conditions and introduce simplifying assumptions regarding the model by design thus not fulfilling the important requirement of a testbed of adequately representing the behavior of the actual physical system. With a HITL testbed, real
hardware is combined with a simulation model contributing to a symbiotic relationship between virtual and physical components of the modeled process.

\begin{figure}[!t]
\centering
\includegraphics[width=0.45\textwidth]{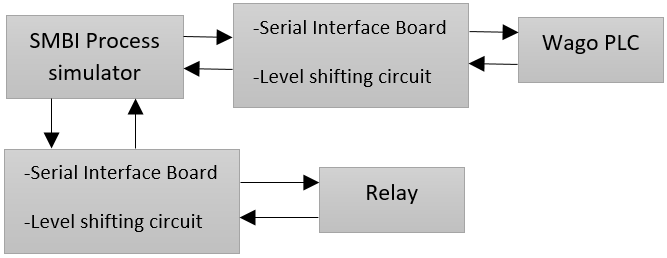}
\caption{HITL testbed including SMBI simulator, Wago PLC, and Relay.}
\label{fig10}
\end{figure}

An experimental testbed (depicted in Figure \ref{fig10}) has been set up to emulate the operation of the SMIB
process. To facilitate this, a Simulink model of the SMIB (run on a PC) has been modified to accommodate a HITL mode of operation wherein a PLC is used sending generation set-point commands to the generator and a protection relay hardware is used controlling the connection of the local load $P_L$ to the SMIB dynamics given in (\ref{eq1}).

The PLC we have chosen for the HITL testbed, as discussed above, is the Wago 750-881 Ethernet Programmable Fieldbus Controller. The Wago PLC features a 32-bit ARM CPU with multitasking capabilities and a real-time clock and runs a Real-Time Operating System (RTOS). The 32KB non-volatile memory hosts configuration files for the PLC, HTML pages, as well as the ladder logic files that get loaded during boot time. Also, the relay is a 750 Feeder Protection Relay, a member of the SR Family of protection relays, is a digital relay for primary protection of transmission lines, feeders, and transformers.

Two serial-interface boards (SIBs), with  analog-to-digital (A/D) and  digital-to-analog (D/A) conversion capabilities and level-shifting circuitry, serves as the interface between the PLC and the relay with the Simulink host PC. Simulink’s serial interface functionality is used to communicate with the SIBs. Packets containing values corresponding to the power generation set-point of the generator are sent from PLC to the SIB over a custom built USB-Serial connection\cite{keliris2016}. Also, the relay open/closed commands are sent from/to PC through the SIB over a USB-Serial connection. The HITL setup includes:

\begin{itemize}
\item Host PC running SMIB model.
\item Serial interface boards (SIBs) and level-shifting (op-amp) array.
\item Wago PLC unit.
\item Relay unit.
\end{itemize}

\vspace*{0.1cm}
\section{Simulation Results} \label{sec3}

\begin{figure}[!t]
\centering
\includegraphics[width=0.45\textwidth]{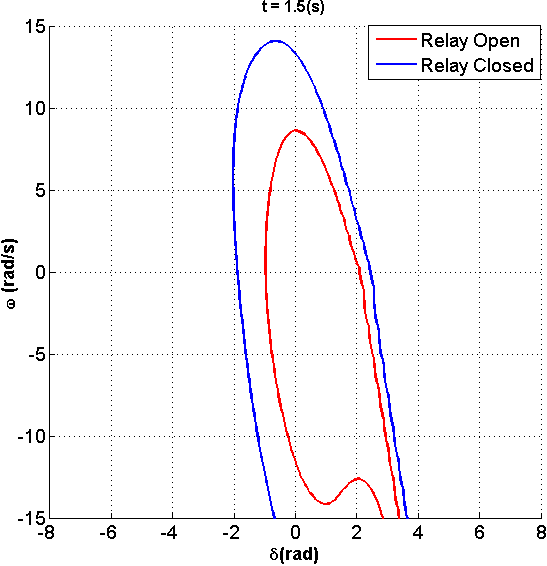}
\caption{Stability region after 1.5 second.}
\label{fig1}
\end{figure}

\begin{figure}[!t]
\centering
\includegraphics[width=0.45\textwidth]{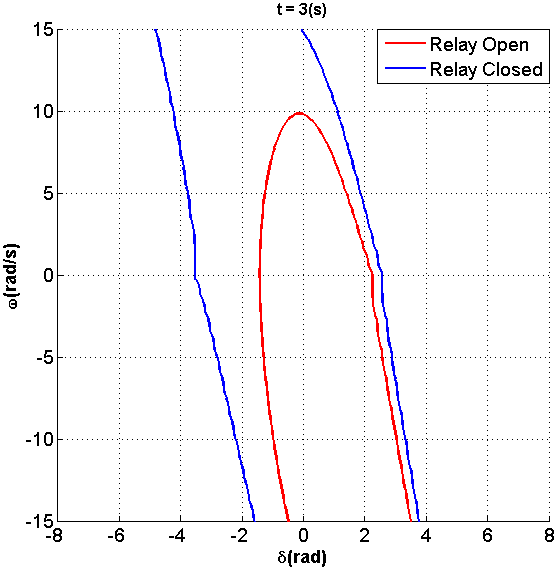}
\caption{Stability region after 3 seconds.}
\label{fig2}
\end{figure}

\begin{figure}[!t]
\centering
\includegraphics[width=0.45\textwidth]{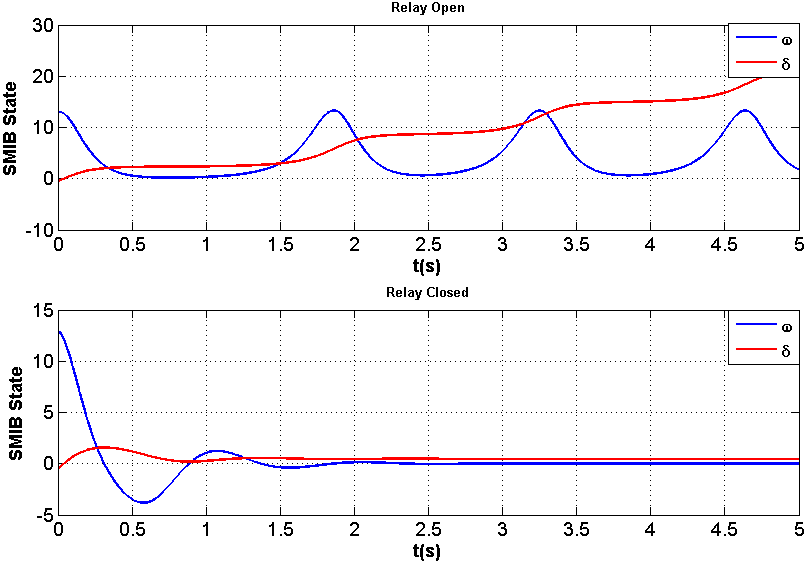}
\caption{Transient response of the system state under different relay statuses.}
\label{fig3}
\end{figure}

In this section, the reachability concept is utilized to answer some security concerns about the SMIB transient stability. First, it is assumed that the attacker only has access to the relay connected to the local load in Figure \ref{fig0}. The objective is to find the stability region of the system when the relay is open/closed. For the system defined in (\ref{eq1}), given $M=0.026 \frac{s^2}{rad}$, $D= 0.12 $, $P_m=1 \ p.u$, $P_G=1.35\  p.u$, and $P_L=0.4\  p.u$, the stability region is computed based on the definition of $Reach(t,S)$, given the target set as a small ball around the stable equilibrium point for the case relay is open/closed. Figures \ref{fig1} and \ref{fig2} show evolution of the stability region after 1.5 and 3 seconds respectively. After $t=3(s)$, the stability region does not change. So, the Points inside the boundaries in Figure \ref{fig2} determine the set of initial conditions that the system will enter target set during $\tau \in [0, 3]$.  The attacker can shrink the stability region of the system by taking control of the relay. Given the initial conditions $[-0.5, 13]^T$, figure \ref{fig3} demonstrates the system trajectory with different relay statuses. These initial conditions are out of stability region for the case relay is open. Therefore, the states do not return to stable equilibrium as shown in figure \ref{fig3}.

\begin{figure}[!t]
\centering
\includegraphics[width=0.45\textwidth]{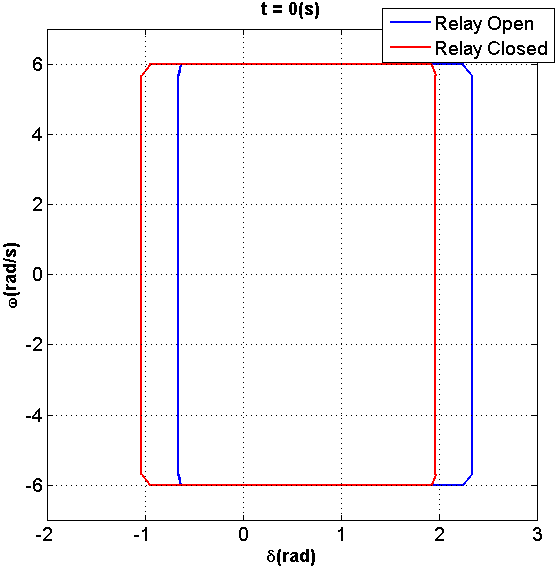}
\caption{Safe set at t=0.}
\label{fig4}
\end{figure}

\begin{figure}[!t]
\centering
\includegraphics[width=0.45\textwidth]{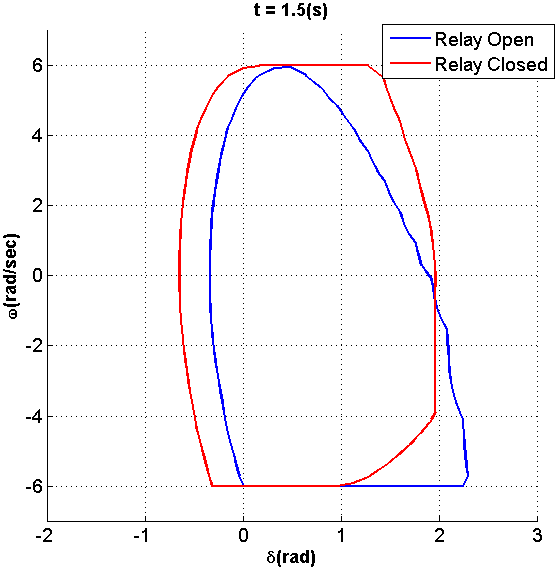}
\caption{Invariant set after 1.5 second.}
\label{fig5}
\end{figure}

\begin{figure}[!t]
\centering
\includegraphics[width=0.45\textwidth]{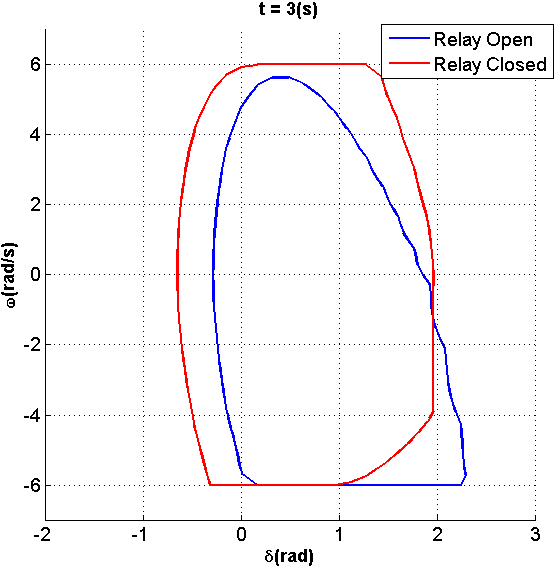}
\caption{Invariant set after 3 second.}
\label{fig6}
\end{figure}

Next, we would like to find the minimum attack signal that keeps the system state from entering the safe set defined in  (\ref{eq0}). Given the definition of the invariant set, if $x \in Inv(0,S)$ the system will never leave the safe set over interval $[0, T]$. We would like to find the set of initial starting points which system state trajectories do not leave the safe set for any attacks policy with a bounded attack signals. it is assumed that the PLC sends a set of control commands to the actuators, e.g., generation set-point to the generator that specifies how much power the generator should generate. Therefore, the attacker can add an arbitrary signal $d$ to the system dynamics by spoofing the PLC control commands to the generator. Figures \ref{fig4} to \ref{fig6} show evolution of the invariant set as time goes when the attacker can inject any arbitrary signal $|d| \leq 0.2$. After $t=3(s)$, the invariant set does not change. Consequently, the intersection of the invariant sets for the relay open/closed status under PLC attack at $t=3(s)$ builds the set of resilient initial points for coordinated attack on relay and PLC. It means that the system will not leave the safe set under any switching policy on relay connected to $P_L$ and any attack signal $|d| \leq 0.2$.

\vspace*{0.1cm}
\section{Experimental Results} \label{sec4}

\begin{figure}[!t]
\centering
\includegraphics[width=0.45\textwidth]{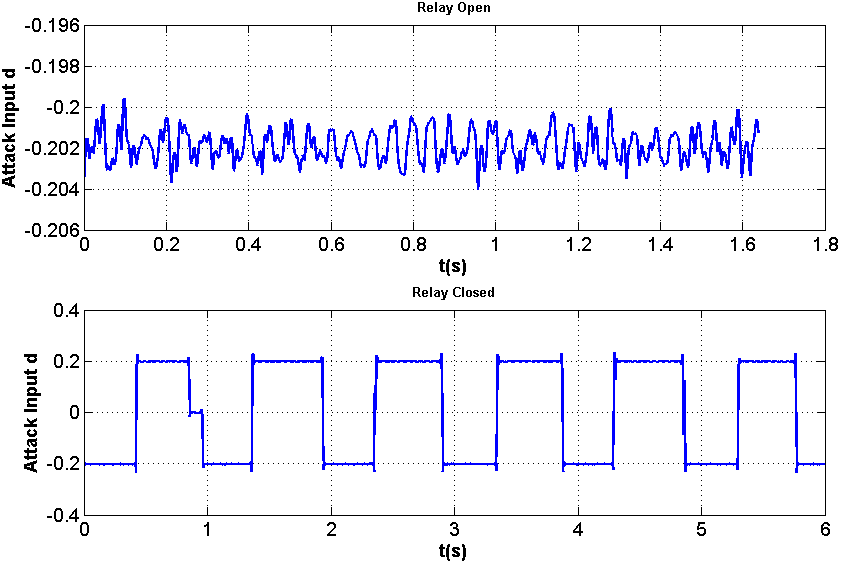}
\caption{PLC optimal attack signal.}
\label{fig7}
\end{figure}

\begin{figure}[!t]
\centering
\includegraphics[width=0.45\textwidth]{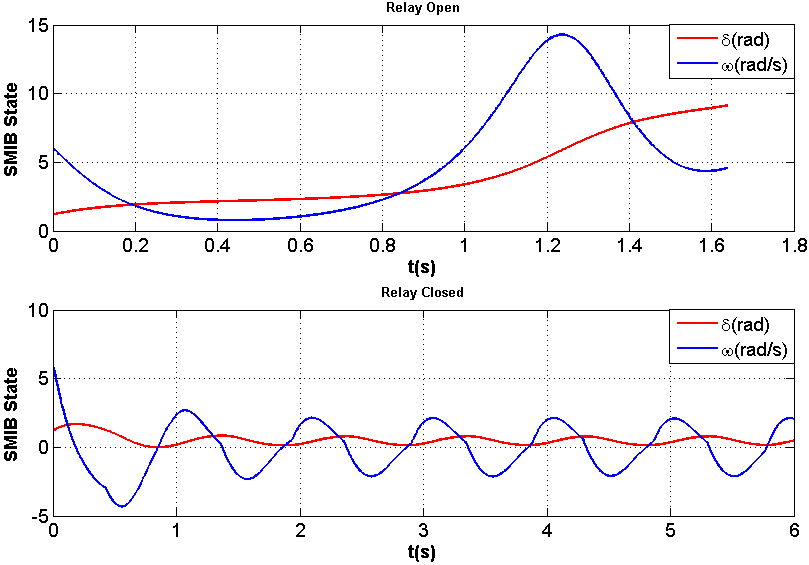}
\caption{System state trajectories under the attack signal $|d|\leq 0.2$.}
\label{fig8}
\end{figure}

For the HITL testbed, the Wago PLC generates the attack signal for Simulink model of the SBIM system. Given (\ref{eq4}), the optimal attack input generated by PLC to keep the system out of the safe set is shown in figure \ref{fig7}, and figure \ref{fig8} represents the state trajectories corresponding to this attack signal. The initial conditions for trajectories shown in figure \ref{fig8} are $[1.2, 6]^T$. The point $[1.2,  6]^T$ is outside the invariant set related to relay open status and inside the invariant set related to relay closed status as shown in figures \ref{fig5} and \ref{fig6}. Consequently, the state trajectory will go outside of the safe set under the optimal attack signal when the relay is open while state trajectory will not leave the safe set even under the optimal attack signal when the relay is closed. These experimental results show that the coordinated attack on PLC and relay gives more flexibility to the attacker to deteriorate system performance.

By performing a series of invariant set computation for different attack signal bounds, it is concluded that the invariant set is empty under attack signal $|d|\geq 0.4$ and $|d|\geq 0.6$ as the relay is open and closed respectively. Empty invariant set means that there is at least one attack signal that can drive the system state out of the safe set and this attack signal can be generated based on (\ref{eq4}).   

\begin{figure}[!t]
\centering
\includegraphics[width=0.45\textwidth]{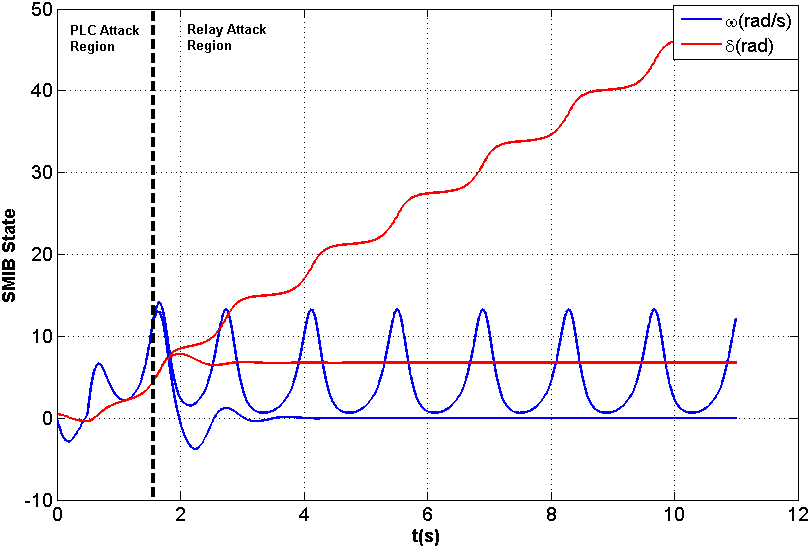}
\caption{System state trajectories under the coordinated attack.}
\label{fig9}
\end{figure}

The attacker does not need to inject an attack signal continuously in order to make the system unstable. Figure \ref{fig9} shows the coordinated attack on PLC and relay. The attacker applies the attack signal till $t=1.93 s$ to keep the system state out of stability region computed in figure \ref{fig2} and after that, he does not apply any signal through PLC and he just makes the relay open to destabilize the system state. As shown in figure \ref{fig9}, note that the system states return to the stability point if the relay remains closed.   

\vspace*{0.1cm}
\section{Conclusion}  \label{sec5}
 
An integrated time-synchronized HITL simulator for the SMBI power system has been developed for experimental demonstration of the effects of the coordinated attacks on the PLC and the relay. It has shown that the attacker can carefully design his attack policy based on reachability analysis on the power system in order to destabilize the generator frequency. Based on the computation of the reachability set, the attacker obtains information about the set of initial conditions that the system states converge to the stable equilibrium. By applying a coordinated attack on the PLC and the relay, the attacker can destabilize the system frequency in a shorter time with respect to the case that the attacker has only control over the PLC commands. Also, the robustness of the systems with respect to the different level of the attack signals was analyzed by leveraging the reachability tools.
\bibliographystyle{IEEEtran}
\bibliography{reach}

\end{document}